# Non-Parametric Estimation of Multiple Periodic Components in Turkey's Electricity Consumption

Jie Yao*

jyao@albany.edu

Edward Valachovic*

*Department of Epidemiology and Biostatistics, School of Public Health, University at Albany, State University of New York, One University Place, Rensselaer, NY 12144

**Abstract**

Electric generation and consumption are an essential component of contemporary living, influencing diverse facets of our daily routines, convenience, and economic progress. There is a high demand for characterizing the periodic pattern of electricity consumption. This study uses the Variable Bandpass Periodic Block Bootstrap (VBPBB) to explore the presence and attributes of periodically correlated (PC) components including daily, weekly, and annual patterns in Turkey's electricity consumption. VBPBB employs a bandpass filter aligned to retain the frequency of a PC component and eliminating interference from other components. This leads to a significant reduction in the size of bootstrapped confidence intervals. Furthermore, other PC bootstrap methods preserve one but not multiple periodically correlated (MPC) components, resulting in superior performance compared to other methods by providing a more precise estimation of the sampling distribution for the desired characteristics. The study of the periodic means of Turkey's electricity consumption using VBPBB is presented and compared with outcomes from alternative bootstrapping approaches. These findings offer significant evidence supporting the existence of daily, weekly, and annual PC patterns, along with information on their timing and confidence intervals for their effects. This information is valuable for enhancing predictions and preparations for future responses to electricity consumption.

## Introduction

Electricity has become a fundamental part of modern life, impacting various aspects of our daily activities, comfort, and economic development. Improving the living conditions, as well as achieving economic and industrial development, is inconceivable without the presence of electricity. It has now evolved into an economic benchmark that delineates a nation's progress or regression.[1] In the present day, electricity consumption stands out as a pivotal metric not only for burgeoning governments, but also for developed countries. Electricity consumption forecasting holds significant potential benefits across various sectors, including energy management, urban planning, and resource allocation.[2,3,4] Electricity consumption forecasting allows utility companies to anticipate demand fluctuations, enabling them to allocate resources more efficiently. By anticipating electricity demand, they can optimize the use of resources such as fossil fuels, reducing greenhouse gas emissions associated with their generation. Electricity markets rely heavily on accurate demand forecasts for efficient trading and market operations. Market participants use these forecasts to make informed decisions regarding energy trading, pricing, and risk management. During extreme weather events or other emergencies, accurate electricity consumption forecasts can help utilities and authorities better prepare for increased demand or potential disruptions to the grid, ensuring a more effective response to emergencies. Precise predictions of electricity consumption play a vital role for policymakers in shaping effective electricity supply policies.

Turkey holds significance in energy policy due to its consumption levels and strategic location as a crucial transit point between Asia and Europe.[5] Similar to other developing nations, Turkey confronts a continuously rising demand for electricity. For instance, from 1980 to 2000, the average annual growth rate of Turkey's total electricity consumption was 8.1%, whereas the real GDP experienced an average annual growth of around 4.4% during the same period.[6] Additionally, per capita electricity consumption has consistently increased from 459 kWh in 1980 to 3,106 kWh in 2023.[7] Ensuring an ample supply of electricity is crucial for meeting the escalating electricity demand and, consequently, for sustaining economic growth in Turkey. Thus, better evidence and characterization of the seasonality of electricity consumption provides vital statistics for the future supply of electricity. Additionally, although this research focuses on Turkey, it can serve as a model applicable to various national, regional, and local contexts, as well as across different fields such as environmental sustainability and resource optimization.

The electricity consumption data is a periodically correlated (PC) time series, which is a time series with a period $p$ exhibiting a strong correlation between data points that are $p$ time points apart. This study examines the seasonal component of electricity consumption in Turkey through a unique non-parametric resampling technique known as the Variable Bandpass Periodic Block Bootstrap (VBPBB).[8,9] This method generates confidence intervals for both seasonal and other periodic means. Bootstrapping, a resampling method, assesses accuracy by randomly sampling with replacement, allowing estimation of bias, variance, confidence intervals, and prediction error for sample estimates. It's particularly useful for time series data, where successive observations are ordered chronologically. However, standard bootstrapping may disrupt correlations between data points. Specialized methods, like block bootstrapping, are necessary to maintain correlation structures in spatio-temporal data. More on time series, including definitions and examples, can be found in Wei.[10]

Although bootstrapping enables the estimation of a statistic's sampling distribution by resampling the original data with replacement, and block bootstrapping serves as a model-free resampling strategy for correlated time series data, neither method succeeds in maintaining correlations within PC time series. Politis introduced a seasonal block bootstrap method where the block length is constrained to be a multiple of the period $p$.[11] In this approach, regardless of the starting point of a block within the cycle, all other blocks in the resample will also commence and conclude at a common step in the cycle of period $p$. This ensures the preservation of correlation structures between data points that are $p$ points apart in time series data. Other block bootstrapping techniques for PC time series include those outlined by Chan et al. and the general seasonal block bootstrap (GSBB) proposed by Dudek et al.[12,13] While the block bootstrap methods mentioned above, collectively referred to here as periodic block bootstrap (PBB) methods, aim to maintain the correlation structure of the PC component, these approaches also bootstrap any additional components, including the noise present in the time series. Notably, even modest levels of noise can lead to widened bootstrapped confidence intervals for measures like the PC component periodic mean beyond the specified confidence interval level.

This study employs the innovative VBPBB. VBPBB utilizes a bandpass filter centered around the frequency of a PC, allowing the preservation of time series variation occurring at or near the targeted component frequency.[8] By segregating and filtering multiple periodically correlated (MPC) components in a time series based on its spectral density according to the frequencies of individual PC components, a collection of component PC time series, each characterized by a distinct PC structure, would be generated. Subsequently, each of these component PC time series, featuring a singular PC structure, could undergo individual block bootstrapping with an appropriate block size to maintain the correlation structure specific to each. VBPBB produces smaller confidence interval (CI) sizes for the periodic mean compared to other PBB methods, such as the GSBB.

A PC component may not adhere strictly to the characteristics of a perfect sinusoidal wave, functioning exclusively at a single principal period or frequency. When a PC component with a period of $p$ or a frequency of $1/p$, referred to as the fundamental frequency, exhibits a pattern divergent from a sinusoidal wave, it frequently displays variations at positive integer multiples of the period, $kp$ for $k \in \mathbb{Z}^+$ or frequency $(1/kp)$, denoted as the $k^{th}$ harmonics of the fundamental frequency. Due to this, this study investigates these harmonics alongside the fundamental seasonal frequency pattern and other PC components of electricity consumption.

In this study, we employ both the GSBB and VBPBB to investigate the periodic patterns in Turkey's electricity consumption. 95% CI bands are constructed based on the 95% CI for the periodic mean electricity consumption at each time point over the study period. A comparison is made between the methods regarding the CI sizes for the periodic means, and significant PC components are identified and quantified. VBPBB outperformed GSBB and identified significant PC components in energy consumption that were not identifiable using the GSBB.

**Methods**

Data Source and Analysis

This analysis utilizes annual time series data spanning from September 6th, 2013, to September 6th, 2023, focusing on industrial electricity consumption measured in kilowatt-hours (KWh) seen in Figure 1. Nominal industrial electricity consumption data were extracted from Energy Exchange Istanbul (EXIST).[14] The electricity consumption data were recorded hourly, enabling investigation of the daily, weekly, monthly, and annual frequency, in addition to their harmonics, based on the periodograms of components. As previously mentioned, these patterns might exhibit harmonics. If we consider the daily PC component as an example, which operates on a 24-hour cycle, the harmonics of the PC occurred at frequency of $1/(24*2), 1/(24*3), 1/(24*4)$, and so on. R 4.1.0 was used to perform all analyses of this study.[15]

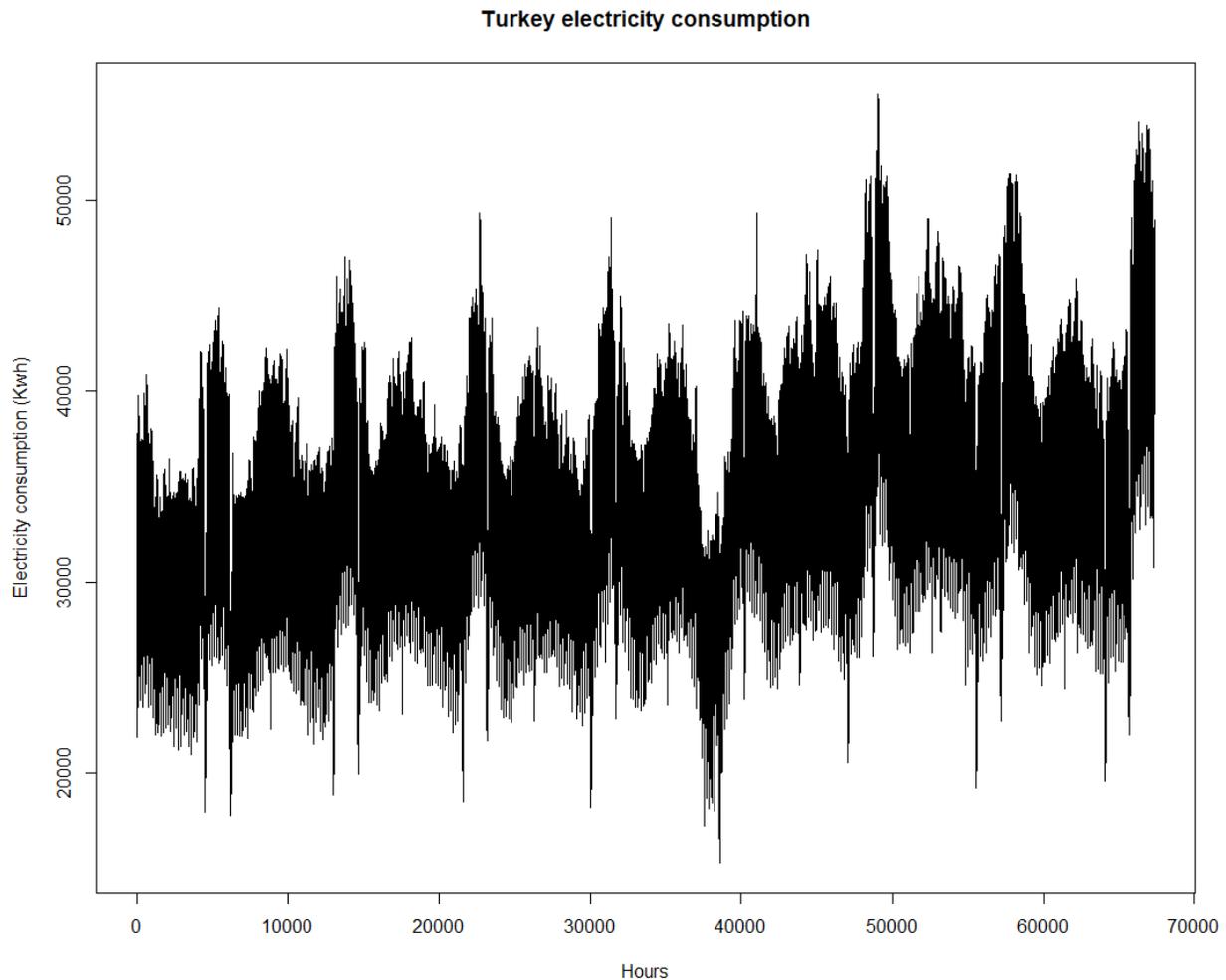

**Figure 1**: Turkey electricity consumption time series.

PBB Approach

For comparative analysis, this study adopts a PBB approach. In contrast to the VBPBB utilized in this work, the PBB approach does not employ a bandpass filter on the PC time series beforehand. The GSBB used here block bootstraps the original electricity consumption time series, preserving the correlation structure of one PC component. This is accomplished by setting the block length to the period $p$, the period of the desired PC component. However, this approach also bootstraps any additional components, including the noise component, which may unnecessarily expand the bootstrapped CIs for statistics such as the periodic mean of the PC component beyond the intended confidence interval level.

VBPBB Approach

The VBPBB bootstrap approach aims to selectively resample a PC component time series to maintain the correlation structure of that specific PC component. This is done without resampling other unrelated components, such as noise or a linear trend, which could unnecessarily increase bootstrap variability—a limitation of the PBB approach. Unlike the PBB method, the VBPBB solution does not operate in the time domain where components interact. Instead, it is initiated in the frequency domain, leveraging the mathematical fact that components of different frequencies are uncorrelated.

The VBPBB method employs bandpass filters to isolate a PC time series based on its spectral density. It separates a narrow band of frequencies, including the PC component frequency, from all others. This results in a new PC time series that is predominantly composed of the desired PC component while retaining the correlation structure. Subsequently, the new PC time series undergoes block bootstrapping with a design similar to that described in the PBB approach, here GSBB, using an appropriate block size to preserve the correlation structure. The VBPBB approach effectively maintains the desired correlation structures within a PC time series while significantly reducing the bootstrapping of uncorrelated components.

The VBPBB employs a bandpass filter to segregate and filter the PC time series, allowing only a narrow band of frequencies around the corresponding frequency of the PC component to pass through. Frequencies outside this band, referred to as the stopband, are attenuated. The bandpass filter used for this purpose is the Kolmogorov-Zurbenko Fourier Transform (KZFT) filter, an extension of iterated moving averages or Kolmogorov-Zurbenko (KZ) filters, initially developed by Zurbenko.[16] KZ filters and their extensions, as described in Yang and Zurbenko, can selectively separate portions of the frequency domain to exclude interfering frequencies.[17] Frequency separation techniques have been developed for multivariate analysis on component factors in Valachovic and Zurbenko's work.[18]

The KZ filter is an iteration of a basic central moving average as defined in Zurbenko.[16] Where the KZ filter is a low pass filter, strongly filtering signals of a frequency at or above the frequency equivalent to $1/m$, the related Kolmogorov-Zurbenko Fourier Transform (KZFT) filter is a band pass filter. In this study, the R statistical software's KZFT function from the KZA package is utilized.[19] Where the KZ filter is symmetric around the zero frequency, the KZFT is a symmetric band pass filter around frequency ν. Practical use of the KZFT filter is similar to the KZ filter since it can be produced in statistical software. The energy transfer function of the KZFT filter centered at a frequency ν with arguments m, a positive odd integer for the filter window length, and k, a positive integer, for the number of iterations is given below:

$$KZFT_{m,k,\nu}(X(t)) = \sum_{u=-\frac{k(m-1)}{2}}^{\frac{k(m-1)}{2}} \frac{a_u^{m,k}}{m^k} e^{-i2m\nu u} X(t+u) \quad (1)$$

The coefficients $a_u^{m,k}$ are the polynomial coefficients from:

$$\sum_{r=0}^{k(m-1)} z^r a_{r-k(m-1)/2}^{m,k} = (1 + z + \cdots + z^{m-1})^k$$

In contrast to the PBB approach, which relies solely on the PC component period for bootstrapping, the VBPBB introduces the need for additional arguments that influence the bandpass filters. This research explores the application of VBPBB to individually bandpass filter each PC component from the electricity consumption time series. VBPBB achieves this by centering a KZFT filter at the frequency corresponding to each PC component. Therefore, for each PC component, a KZFT filter should have ν set to the frequency of that specific PC component. To selectively pass or retain only one PC component in each KZFT bandpass filter, the width of passed frequencies should be set to no more than halfway between the minimum bandwidth among all frequencies to be filtered.

For every pair of two PC components with periods $p_1$ and $p_2$ and corresponding frequencies $\nu_1 = 1/p_1$ and $\nu_2 = 1/p_2$, KZFT filters are centered at $\nu_1$ and $\nu_2$. The number of iterations for the KZFT filters are set to $k = 1$ since higher iterations require increasing larger quantities of data to fully apply the KZFT filter and only serve to further suppress the frequencies outside of the bandpass region. For the purposes of this work, one iteration is adequate to sufficiently suppress frequencies outside of the bandpass. The argument $m$ is derived from the closest odd integer larger than $2/|\nu_1 - \nu_2| = 2(p_1 p_2)/|p_1 - p_2|$. For an MPC time series comprising $m$ PC components, where $i$ takes values from 1 to $m$, the KZFT employed to bandpass each individual PC component follows the arguments: $\nu = \nu_i$, $k = 1$, and $m$ set to the maximum value that ensures exclusion of all other frequencies associated with different PC components, as detailed earlier. In this work the closest two frequencies are the annual cycle and its harmonics, which is m = (365*48) = 8760 for each application of the KZFT filter.

Following the separation of PC components, the VBPBB approach employs block bootstrapping on the individual PC component time series rather than the original PC time series. The same bootstrapping procedure, as explained earlier for the PBB approach, is then applied to these component time series. This iterative process is carried out for a substantial number, $B$, of resamples, where $B$ is set to 1,000 in this study. Subsequently, the 0.975 and 0.025 quantiles of the $B$ bootstrapped time series resample periodic means are computed, resulting in a 95% confidence interval for the periodic mean at each time point. This compilation forms a VBPBB 95% CI band, providing an interval estimate for the periodic mean across the specified time interval. To test the null hypothesis that the periodic mean variation of PC component is zero, we calculate the maximum of the lower 95% CI band and the minimum of the upper 95% CI band. If the 95% CI band excludes a flat line representing no variation at that PC component frequency, the PC component is statistically significant. Otherwise, we cannot exclude the possibility that there may be no periodic mean variation at that PC component frequency. Additionally, the correlation squared, or coefficient of determination, is used to explain how much of the total variability in the original time series was contributed by the PC component.

**Results**

Table 1 presents the 95% CI band for the 8 PC components' mean variation of the adjusted electricity consumption. From the table, the 95% CI range of the daily cycle periodic variation high value is (3152.596, 5993.240) kWh and the range of 95% CI range for daily cycle periodic variation low value is (-6069.338, -3070.536) kWh. As 0 is out of the 95% CI bounds, the daily cycle is significant. Similarly, the 95% CI bounds of daily cycle' second through fourth harmonics do not include 0, so all daily cycle components are significant. For weekly cycle and annual cycle components, the 95% CI bounds also do not contain 0. Thus, all the 8 PC components are significant.

**Table1. The 95% CI band for the 8 periodically correlated components' means variation.**

| Variable | Frequency (1/hours) | Lower Range 95% CI | Upper Range 95% CI |
|---|---|---|---|
| Daily cycle | $\frac{1}{24}$ | **(-6069.338, -3070.536)*** | **(3152.596, 5993.240)*** |
| Daily cycle (The second harmonic) | $\frac{2}{24}$ | **(-2423.410, -1037.955)*** | **(1009.043, 2450.511)*** |
| Daily cycle (The third harmonic) | $\frac{3}{24}$ | **(-933.881, -41.837)*** | **(62.16146, 905.8671)*** |
| Daily cycle (The fourth harmonic) | $\frac{4}{24}$ | **(-1019.372, -39.112)*** | **(67.741, 977.637)*** |
| Weekly cycle | $\frac{1}{168}$ | **(-3088.389, -1158.549)*** | **(1174.300, 3055.381)*** |
| Weekly cycle (The second harmonic) | $\frac{2}{168}$ | **(-2005.291, -725.371)*** | **(841.586, 2019.013)*** |
| Weekly cycle (The third harmonic) | $\frac{3}{168}$ | **(-1200.173, -473.250)*** | **(497.555, 1211.775)*** |
| Annual cycle | $\frac{1}{8736}$ | (-7690.450, 245.8552) | (-251.636, 11468.650) |
| Annual cycle (The second harmonic) | $\frac{2}{8736}$ | **(-7657.379, -2382.784)*** | **(2272.894, 9047.811)*** |
| Annual cycle (The third harmonic) | $\frac{3}{8736}$ | (-2318.211, 141.115) | (-202.8373, 2617.901) |

**\* P < 0.05**

Figure 2 depicts the bootstrapped 95% CI band for the annual (The second harmonic) pattern, showcasing PBB in red and VBPBB in blue over typical 365-day cycles. Within the CI band, the median PBB CI size is 1.64 times larger than that of VBPBB. The 95% CI band of VBPBB offers ample evidence supporting the significance of the second harmonic of annual cycle, leading to the rejection of the null hypothesis that the annual cycle of the seasonal mean variation is zero. In contrast, PBB does not provide such evidence.

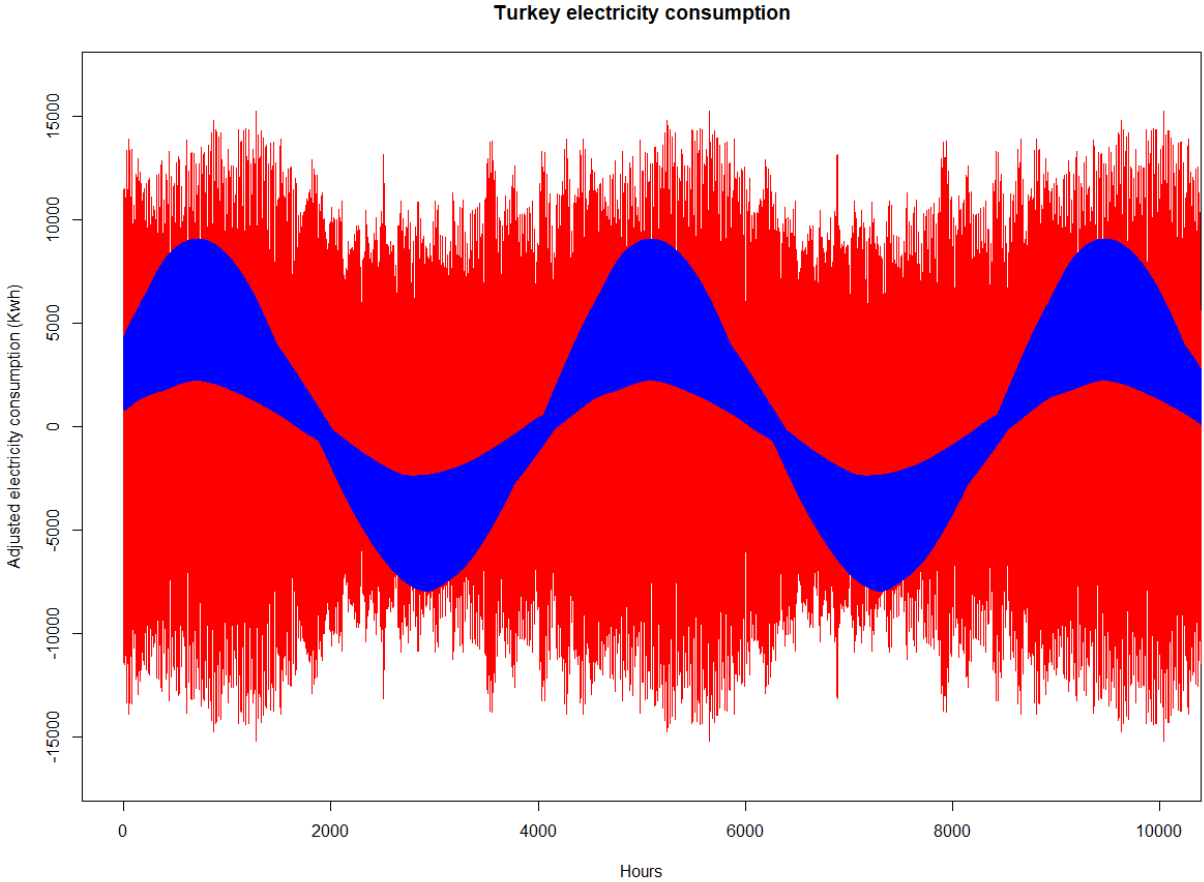

**Figure 2**: 95% CI band for the annual (The second harmonic) mean component variation of the adjusted electricity consumption annual variation with PBB in red and VBPBB in blue.

Figure 3 depicts the bootstrapped 95% CI band for the weekly pattern along with the second harmonic through the third harmonic, showcasing PBB in red and VBPBB in blue over typical 7-day cycles. Within the CI band, the median PBB CI size is 4.61 times larger than that of VBPBB. The 95% CI band of VBPBB again offers evidence supporting the significance of the weekly cycle, leading to the rejection of the null hypothesis that the weekly cycle of the seasonal mean variation is zero. In contrast, PBB does not provide such evidence.

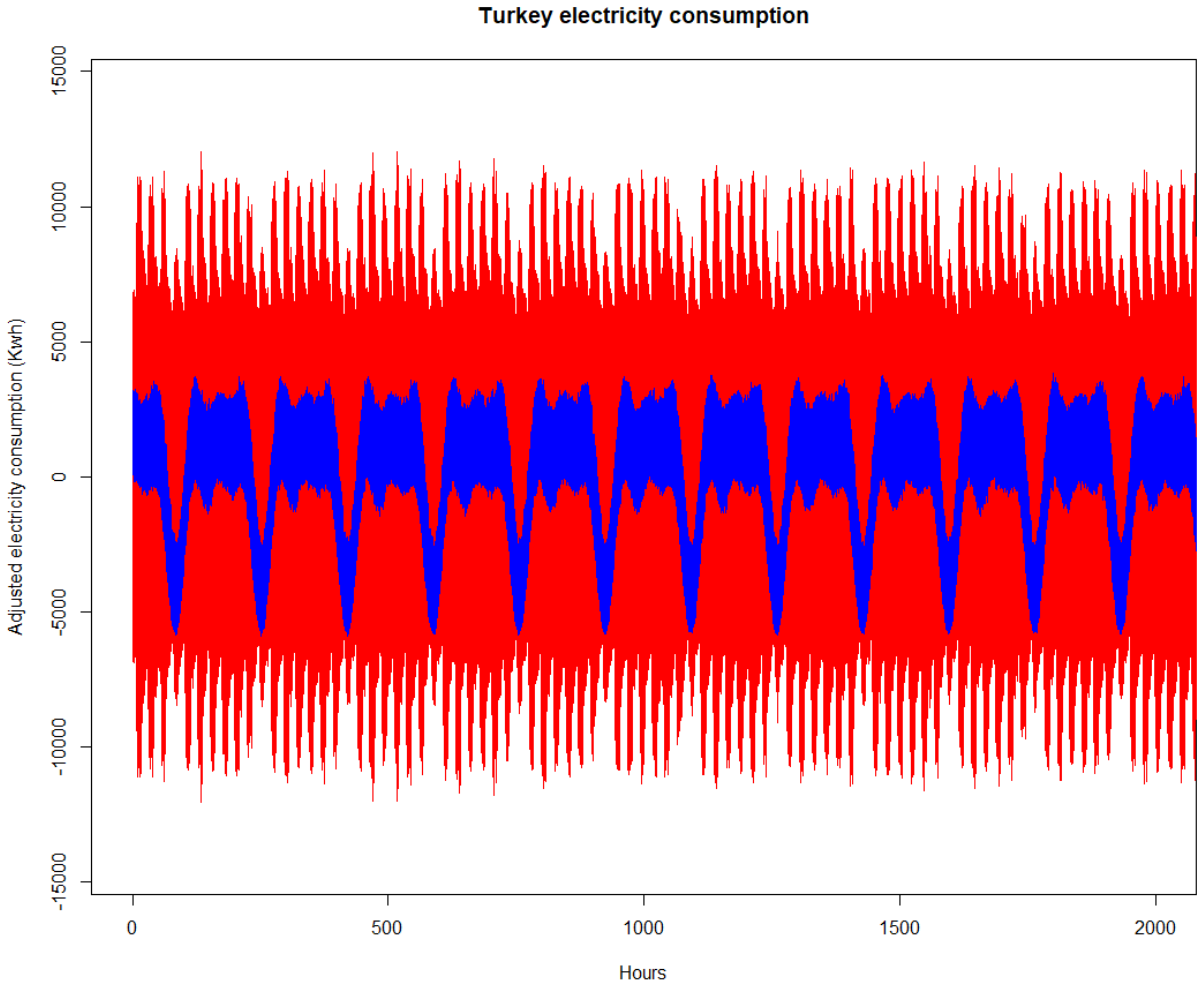

**Figure 3**: 95% CI band for the weekly mean component along with the second harmonic through the third harmonic variation of the adjusted electricity consumption weekly variation with PBB in red and VBPBB in blue.

Figure 4 depicts the bootstrapped 95% CI band for the daily pattern along with the second harmonic through the fourth harmonic, showcasing PBB in red and VBPBB in blue over typical 24-hour cycles. Within the CI band, the median PBB CI size is 3.65 times larger than that of VBPBB. The 95% CI band of VBPBB offers strong evidence supporting the significance of the daily cycle along with its second through fourth harmonics, leading to the rejection of the null hypothesis that the daily cycle of the seasonal mean variation is zero. In contrast, PBB does not provide such evidence.

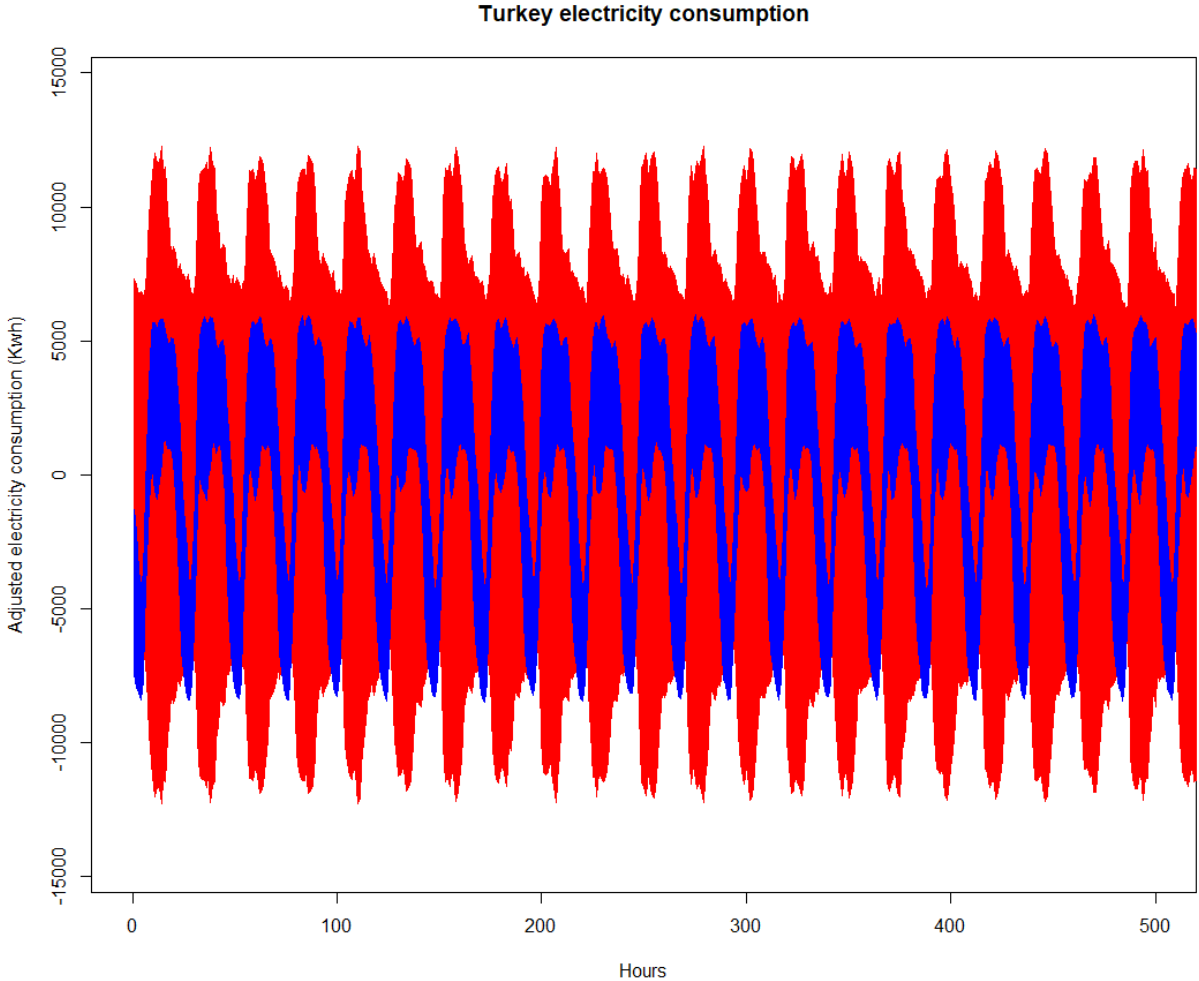

**Figure 4**: 95% CI band for the daily mean component along with the second harmonic through the fourth harmonicvariation of the adjusted electricity consumption daily variation with PBB in red and VBPBB in blue.

Figure 5 illustrates the VBPBB 95% confidence interval (CI) band in blue for the combination of daily component along with its second through fourth harmonics, the weekly component with its second through third harmonics, and the second harmonic of annual component. The PBB CI band that pertains only to the 24-hour variation is noted to be frequently larger than the VBPBB CI band for all combined daily, daily's harmonic, weekly, weekly's harmonic and annual components. The median PBB to the 24-hour variation CI size is 1.58 times larger than that of VBPBB.

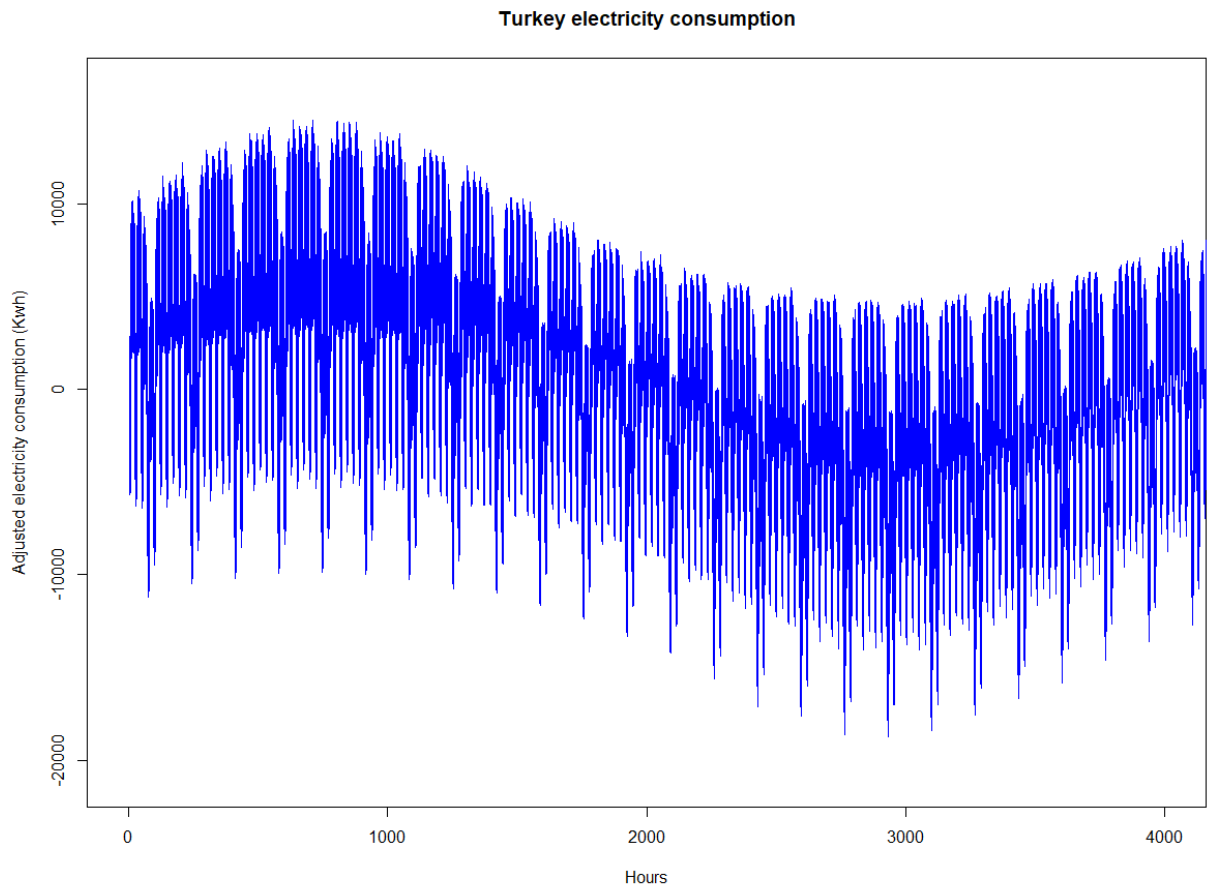

**Figure 5**: The VBPBB 95% CI band for the adjusted electricity consumption periodic mean variation for the combination of significant daily, weekly, and annual cycles and harmonics in blue.

In conclusion, the coefficient of determination, also known as the square of correlation, between the original time series of the electricity consumption and the median of the significant PC, encompassing the daily, daily's harmonic, weekly, weekly's harmonic and annual's second harmonic frequencies, is 0.466 implying the significant PC components explain 46.6% of the variation in electricity consumption.

**Discussion**

VBPBB has identified 8 distinct significant components in the PC analysis of Turkey's electricity consumption. These components operate at different periods or frequencies, encompassing the daily component along with its second through fourth harmonics, the weekly component with its second through third harmonics, and the annual's second harmonic component. On the other hand, PBB did not identify any significant PC components in the data. This investigation is important to provide significant evidence that there were MPC components in electricity consumption. VBPBB emerges as a solution for time series with MPC components, whereas PBB falls short in reflecting the variation of all PC components since it only focuses on one frequency at a time.

To compare the daily mean component variation between VBPBB and PBB, we combined the daily component with all the harmonics in VBPBB and compare the 95% CI band size for each method, The median PBB 95% CI size is 3.65 times larger than that of VBPBB. The correlation squared between the bootstrapped daily PC component of VBPBB, and the original time series is around 0.332. This implies that the daily PC component contributes to roughly 33.2% of the total variability in Turkey's electricity consumption time series. Similarly, we compared the 95% CI weekly mean component variation band size between PBB and VBPBB of the combination of weekly component and its harmonics. The median PBB 95% CI size is 4.61 times larger than that of VBPBB. The correlation squared between the bootstrapped weekly PC component of VBPBB, and the original time series is around 0.101. This implies that the weekly PC component contributes to roughly 10.1% of the total variability in Turkey's electricity consumption time series. Furthermore, we compared the 95% CI annual mean component variation band size between PBB and VBPBB. The median PBB 95% CI size is 1.64 times larger than that of VBPBB. The correlation squared between the bootstrapped annual PC component of VBPBB, and the original time series is around 0.033. This implies that the annual PC component contributes to roughly 3.3% of the total variability in Turkey's electricity consumption time series.

The design of VBPBB exhibits several limitations. The performance of VBPBB is intricately linked to the selection of arguments for bandpass filtration. There are restrictions on which frequencies can be detected and the degree of proximity two frequencies can have while remaining distinguishable through KZFT filters. Since VBPBB methods rely on filtering the original PC time series through the application of moving averages, the choice of moving average arguments in the KZFT filter will affect the final performance of VBPBB. The process of segregating components based on frequency needs to adhere to specific guidelines outlined by Valachovic.[20] Therefore, these results should be considered preliminary, and we recommend additional research which will potentially improve the characterization of the PC components.

We advocate for the application of this innovative VBPBB method to additional electricity consumption datasets, including but not limited to sensible and reliable energy demand forecasts, and energy policy adjustment which can significantly contribute to public health preparation and financing and developing the necessary measures for the sustainable economic growth. As more data becomes accessible, the performance of VBPBB is likely to improve, rendering results more reliable. With additional data, other PC components may gain significance, and the existing confidence interval bands for the PC components of electricity consumption, revealing timing and intensity, can be refined.